\documentclass[fleqn,usenatbib,usedcolumn]{mnras}

\usepackage{array}

\usepackage{enumerate}
\usepackage{aas_macros}

\usepackage{amsmath}
\usepackage{mathtools}
\usepackage{IEEEtrantools}
\usepackage{float}

\usepackage{upgreek}

\usepackage{txfonts}

\usepackage[applemac]{inputenc}
\usepackage[T1]{fontenc} 

\usepackage[english]{babel}
\usepackage{newtxtext}                  % Good fonts
\usepackage[slantedGreek]{newtxmath}    %   "    "   (slanted Greek)

\usepackage{booktabs}
\usepackage{multirow}

\usepackage{savesym}
\savesymbol{iint}
\savesymbol{iiint}
\savesymbol{iiiint}
\savesymbol{idotsint}

\usepackage{amssymb,amsmath}
\usepackage{pifont}% http://ctan.org/pkg/pifont

\usepackage{amstext}
\usepackage{graphicx}
\usepackage{caption}
\usepackage{subcaption}
\usepackage[np,noautolanguage]{numprint}

\usepackage{bigstrut}

\usepackage{scalefnt}
\usepackage{letltxmacro}
\usepackage{color}

\usepackage{mathrsfs}

\hypersetup{breaklinks, colorlinks, linktocpage=true, linkcolor=blue, citecolor=blue, filecolor=black, urlcolor=blue}
\usepackage{relsize}

\usepackage{etoolbox}
\makeatletter
 
\patchcmd{\NAT@citex}
  {\@citea\NAT@hyper@{%
     \NAT@nmfmt{\NAT@nm}%
     \hyper@natlinkbreak{\NAT@aysep\NAT@spacechar}{\@citeb\@extra@b@citeb}%
     \NAT@date}}
  {\@citea\NAT@nmfmt{\NAT@nm}%
   \NAT@aysep\NAT@spacechar\NAT@hyper@{\NAT@date}}{}{}

\patchcmd{\NAT@citex}
  {\@citea\NAT@hyper@{%
     \NAT@nmfmt{\NAT@nm}%
     \hyper@natlinkbreak{\NAT@spacechar\NAT@@open\if*#1*\else#1\NAT@spacechar\fi}%
       {\@citeb\@extra@b@citeb}%
     \NAT@date}}
  {\@citea\NAT@nmfmt{\NAT@nm}%
   \NAT@spacechar\NAT@@open\if*#1*\else#1\NAT@spacechar\fi\NAT@hyper@{\NAT@date}}
  {}{}

\makeatother
\bibpunct{(}{)}{;}{a}{}{,} % to follow the A&A style

\usepackage{threeparttable}

\numberwithin{equation}{section}

 \usepackage{upgreek}
\usepackage{tabularx}

\LetLtxMacro{\oldtextsc}{\textsc}
\renewcommand{\textsc}[1]{\oldtextsc{\scalefont{1.2}#1}}

\newcommand{\affilSpace}{-5pt}

\newcommand{\SBID}[1]{\hbox{SB #1}}

\newcommand{\txn}[1]{\textnormal{#1}}
\newcommand{\range}[3]{\hbox{$#1 \sim #2 \,\txn{--}\, #3$}}
\newcommand{\rangeto}[3]{\hbox{$#1 \sim #2 \;\txn{to}\, #3$}}

\newcommand{\lam}[1]{\hbox{$\,\uplambda#1$}}
\newcommand{\lamlam}[2]{\hbox{$\,\uplambda\uplambda#1,\!#2$}}

\renewcommand{\bmath}[1]{\mbox{ \boldmath $\!#1\!$ \unboldmath}}

\newcommand{\smallerSub}[1]{{\scriptscriptstyle{#1}}}
\newcommand{\smallerSubTxn}[1]{\txn{\scriptsize{#1}}}

\newcommand{\conditional}[2]{\hbox{$\txn{P}(#1 \mid #2)$}}

\newcommand{\thetab}{\hbox{$\bmath{\upTheta}$}}
\newcommand{\Db}{\hbox{$\bmath{D}$}}
\newcommand{\prior}{\hbox{$\pi(\thetab)$}}
\newcommand{\likelihood}{\hbox{$\mathcal{L}(\thetab)$}}

\newcolumntype{L}[1]{>{\raggedright\let\newline\\\arraybackslash\hspace{0pt}}m{#1}}
\newcolumntype{C}[1]{>{\centering\let\newline\\\arraybackslash\hspace{0pt}}m{#1}}
\newcolumntype{R}[1]{>{\raggedleft\let\newline\\\arraybackslash\hspace{0pt}}m{#1}}

\newcommand{\EW}{\hbox{$\txn{EW}$}}
\newcommand{\vs}{\textit{vs}}

\newcommand{\Zsun}{\hbox{$\Z_\odot$}}

\newcommand{\yr}{\hbox{$\txn{yr}$}}

\newcommand{\JWST}{\textit{JWST}}
\newcommand{\HST}{\textit{HST}}

\newcommand{\Spitzer}{\textit{Spitzer}}

\newcommand{\beagle}{\hbox{\textsc{beagle}}}

\newcommand{\cloudy}{\hbox{\textsc{cloudy}}}

\newcommand{\multinest}{\textsc{multinest}}

\newcommand{\tauV}{\hbox{$\hat{\tau}_\smallerSub{V}$}}

\renewcommand{\t}{\hbox{$t$}}

\newcommand{\logt}{\hbox{$\log{(\t/\yr)}$}}

\newcommand{\nH}{\hbox{$n_\smallerSubTxn{H}$}}

\newcommand{\Z}{\hbox{$\txn{Z}$}}
\newcommand{\logZ}{\hbox{$\log(\Z/\Zsun)$}}

\newcommand{\M}{\hbox{$\txn{M}$}}
\newcommand{\logM}{\hbox{$\log(\M/\Msun)$}}

\newcommand{\Msun}{\hbox{$\txn{M}_{\odot}$}}

\newcommand{\Muv}{\hbox{$M_\smallerSubTxn{UV}$}}

\newcommand{\Luv}{\hbox{$L_\smallerSubTxn{UV}$}}
\newcommand{\LuvHII}{\hbox{$L_\smallerSubTxn{UV}^\smallerSubTxn{H\,\textsc{ii}}$}}
\newcommand{\Luvst}{\hbox{$L_\smallerSubTxn{UV}^\star$}}

\newcommand{\fesc}{\hbox{$f_{\mathrm{esc}}$}}

\newcommand{\Us}{\hbox{$U_\smallerSubTxn{S}$}}
\newcommand{\logUs}{\hbox{$\log \Us$}}
\newcommand{\xid}{\hbox{$\upxi_\txn{d}$}}
\newcommand{\Zism}{\hbox{$\Z_\smallerSubTxn{ISM}$}}

\newcommand{\Te}{\hbox{$T_e$}}

\newcommand{\logOH}{\hbox{$12 + \log (\txn{O}/\txn{H})$}}

\newcommand{\CO}{\hbox{$\txn{C}/\txn{O}$}}

\newcommand{\Nion}{\hbox{$\dot{N}_\txn{ion}$}}

\newcommand{\xiion}{\hbox{$\upxi_\txn{ion}$}}
\newcommand{\xiionHII}{\hbox{$ \upxi_\txn{ion}^\smallerSubTxn{H\,\textsc{ii}}$}}
\newcommand{\xiionst}{\hbox{$ \upxi_\txn{ion}^\star$}}

\newcommand{\Hz}{\hbox{H$\upzeta$}}

\newcommand{\Hd}{\hbox{H$\updelta$}}
\newcommand{\Hg}{\hbox{H$\upgamma$}}
\newcommand{\Hb}{\hbox{H$\upbeta$}}
\newcommand{\Ha}{\hbox{H$\upalpha$}}

\newcommand{\HeILam}[1]{\hbox{He\,{\sc i}\lam{#1}}}

\newcommand{\HeII}{\hbox{He\,{\sc ii}\lam{1640}}}
\newcommand{\HeIInoL}{\hbox{He\,{\sc ii}}}
\newcommand{\HeIIopt}{\hbox{He\,{\sc ii}\lam{4686}}}

\newcommand{\OIIIuv}{\mbox{O\,{\sc iii]}\lamlam{1660}{1666}}}
\newcommand{\OI}{\mbox{[O\,{\sc i]}\lam{6363}}}
\newcommand{\OII}{\mbox{[O\,{\sc ii]}\lamlam{3726}{3729}}}

\newcommand{\OIIInoL}{\mbox{[O\,{\sc iii]}}}
\newcommand{\OIII}{\mbox{[O\,{\sc iii]}\lamlam{4959}{5007}}}
\newcommand{\OIIIa}{\mbox{[O\,{\sc iii]}\lam{4959}}}

\newcommand{\OIIIc}{\mbox{[O\,{\sc iii]}\lam{4363}}}

\newcommand{\NII}{\mbox{[N\,{\sc ii]}\lamlam{6548}{6584}}}
\newcommand{\NIInoL}{\mbox{[N\,{\sc ii]}}}

\newcommand{\NIIb}{\mbox{[N\,{\sc ii]}\lam{6584}}}

\newcommand{\SII}{\mbox{[S\,{\sc ii]}\lamlam{6716}{6731}}}
\newcommand{\SIInoL}{\mbox{[S\,{\sc ii]}}}

\newcommand{\Lya}{\hbox{Ly$\alpha$}}

\newcommand{\CIII}{\hbox{[C\,{\sc iii}]\lam1907+C\,{\sc iii}]\lam1909}}
\newcommand{\CIIInoL}{\hbox{C\,{\sc iii}]}}

\newcommand{\CIV}{\mbox{C\,{\sc iv}\lamlam{1548}{1551}}}

\newcommand{\HII}{\mbox{H\,{\sc ii}}}

\usepackage{stfloats}

\title[Physical properties of extreme nearby star-forming regions]{Physical properties and H-ionizing-photon production rates of extreme nearby star-forming regions}
\author[J.~Chevallard]{
Jacopo~Chevallard$^{1}$\thanks{E-mail: jchevall@cosmos.esa.int}\thanks{ESA Research Fellow},
St\'ephane~Charlot$^{2}$,
Peter~Senchyna$^{3}$,
Daniel P. Stark$^{3}$,
\vspace{-8pt} \newauthor \vspace{-8pt}
Alba Vidal-Garc\'ia$^{2}$,
Anna Feltre$^{4}$,
Julia Gutkin$^{2}$,
Tucker Jones$^{5,6}$\thanks{Hubble Fellow},
Ramesh Mainali$^{3}$,
\newauthor
Aida Wofford$^{7}$
\\
\vspace{\affilSpace} $^{1}$Scientific Support Office, Directorate of Science and Robotic Exploration, ESA/ESTEC, Keplerlaan 1, 2201 AZ Noordwijk, The Netherlands\\
\vspace{\affilSpace} $^{2}$Sorbonne Universit\'es, UPMC-CNRS, UMR7095, Institut d'Astrophysique de Paris, F-75014, Paris, France\\
\vspace{\affilSpace} $^{3}$Steward Observatory, University of Arizona, 933 N Cherry Ave, Tucson, AZ 85721 USA \\
\vspace{\affilSpace} $^{4}$Centre de Recherche Astrophysique de Lyon, Universit\'e Lyon 1, 9 Avenue Charles Andr\'e, F-69561 Saint Genis Laval Cedex, France \\
\vspace{\affilSpace} $^{5}$Department of Physics, University of California Davis, 1 Shields Avenue, Davis, CA 95616, USA \\
\vspace{\affilSpace} $^{6}$Institute for Astronomy, University of Hawaii, 2680 Woodlawn Drive, Honolulu, HI 96822, USA \\
\vspace{\affilSpace} $^{7}$Instituto de Astronom\'ia, UNAM, Ensenada, CP 22860, Baja California, Mexico
}

\date{Accepted . Received ; in original form }

\pubyear{2017}

\begin{document}
\label{firstpage}
\pagerange{\pageref{firstpage}--\pageref{lastpage}}
\maketitle

% Abstract
\begin{abstract}

Measurements of the galaxy UV luminosity function at $z \gtrsim 6$ suggest that young stars hosted in low-mass star-forming galaxies produced the bulk of hydrogen-ionizing photons necessary to reionize the intergalactic medium (IGM) by redshift $z\sim6$. Whether star-forming galaxies dominated cosmic reionization, however, also depends on their stellar populations and interstellar medium properties, which set, among other things, the production rate of H-ionizing photons, \xiionst, and the fraction of these escaping into the IGM. 
Given the difficulty of constraining with existing observatories the physical properties of $z \gtrsim 6$ galaxies, in this work we focus on a sample of ten nearby objects showing UV spectral features comparable to those observed at $z\gtrsim 6$. We use the new-generation \beagle\ tool to model the UV-to-optical photometry and UV/optical emission lines of these Local `analogues' of high-redshift galaxies, finding that our relatively simple, yet fully self-consistent, physical model can successfully reproduce the different observables considered. Our galaxies span a broad range of metallicities and are characterised by high ionization parameters, low dust attenuation, and very young stellar populations. Through our analysis, we derive a novel diagnostic of the production rate of H-ionizing photons per unit UV luminosity, \xiionst, based on the equivalent width of the bright \OIII\ line doublet, which does not require measurements of H-recombination lines. This new diagnostic can be used to estimate \xiionst\ from future direct measurements of the \OIII\ line using \JWST/NIRSpec (out to $z\sim9.5$), and by exploiting the contamination by $\Hb+\OIII$ of photometric observations of distant galaxies, for instance from existing  \Spitzer/IRAC data and from future ones with \JWST/NIRCam. 
 
% , paving the way to similar analyses at higher redshifts with \JWST/NIRSpec

\end{abstract}

%TC:ignore

%Keywords

\section{Introduction}\label{sec:intro}

The appearance of the first stars and galaxies in the Universe at the end of the Dark Ages marked the beginning of the epoch of reionization (EoR, e.g. \citealt{Bromm2011}). The EoR appears to have lasted mostly between $z\sim 15$ \citep{Planck2016} and $z\sim6$ \cite[e.g.][]{Fan2006}. Reionization is generally thought to have been mainly driven by young stars in low-mass galaxies \citep[see review by][]{Stark2016}, with minor contributions by active galactic nuclei \citep[e.g.][]{Parsa2017}, X-ray binaries \citep[e.g.][]{Mirabel2011} and Population-III stars \citep[e.g.][]{Wise2014}. The observed steepening of the faint-end slope of the ultraviolet (UV) galaxy luminosity function at redshift \range{z}{4}{8} \citep[e.g.][]{Bouwens2015a,Finkelstein2015} in fact suggests that low-mass galaxies were more abundant in the early Universe than at more recent times, potentially providing enough H-ionizing photons to reionize the IGM \citep[e.g.][]{Robertson2015, Bouwens2015b}. However, this interpretation is subject to major uncertainties: current data do not allow any accurate determination of the abundances of low-mass galaxies during the EoR, because of degeneracies among the luminosity function parameters \citep[e.g.][]{Finkelstein2015} and the difficulty of probing magnitudes fainter than $\Muv \sim-17$ (except in lensed fields, e.g. \citealt{Atek2015a, Atek2015b, Ishigaki2018}). Another major source of uncertainty is the relation between the (non-ionizing) UV-continuum emission of EoR galaxies, their production rate of H-ionizing photons, \Nion, and the number of these photons escaping from the galaxies, \fesc. Direct measurements of the rate of ionizing photons escaping from galaxies (the product $\fesc \times \Nion$) are only possible at lower redshifts \citep[e.g.][]{Izotov2016, Vanzella2016}, since any H-ionizing photon emitted by EoR galaxies would be absorbed by neutral H along the line of sight. Indirect diagnostics on $\fesc$ have been proposed, for instance based on the relation between H-Balmer lines and UV-continuum emission \citep{Zackrisson2013} and on the \Lya-line profile \citep{Verhamme2015, Verhamme2017}, but have not yet been tested on EoR galaxies.

%the contribution by star-forming galaxies to cosmic reionization depends on several other largely unconstrained quantities. These include the minimum absolute UV magnitude of low-mass galaxies, production rate of H-ionizing photons ($\Nion$), fraction of these escaping into the IGM (\fesc), and spatial variations of density and temperature in the IGM.

In this paper, we focus on the production rate of H-ionizing photons per unit intrinsic monochromatic UV (1500\,\AA) luminosity, $\xiionst=\Nion/\Luvst$. This quantity depends only on the properties of the stellar populations emitting the ionizing photons, i.e. mainly age and metallicity. Analytical reionization models often assume a fixed value for \xiionst, typically in the range \range{\log(\xiionst/\txn{erg}^{-1}\,\txn{Hz})}{25.2}{25.3} \citep[e.g.][]{Kuhlen2012, Robertson2013}, although the dependence of \xiionst\ on stellar population properties would introduce galaxy-to-galaxy variations (both stochastic and systematic). Here, we use the new-generation spectral analysis tool \beagle\ \citep{Chevallard2016} to measure the  production rate of H-ionizing photons of ten extreme nearby star-forming regions found by \citet[][hereafter S17]{Senchyna2017} to have UV spectra similar to those of $z\sim6$--7 star-forming galaxies. Our simultaneous modelling of several UV and optical emission lines also provide valuable constraints on the properties of the ionizing stars, photoionized gas and attenuation by dust. Beyond the demonstration that the stellar population and photoionization models incorporated into \beagle\ offer accurate fits to the data, this analysis allows us to uncover and calibrate a relation between \xiionst\ and equivalent width of ($\Hb+$) \OIII. This relation extends from canonical \xiionst\ values at low line equivalent widths up to nearly an order or magnitude larger for the most extreme galaxies. 

%Direct studies of the physical conditions of gas and of stellar population in galaxies at $z\gtrsim6$ require measuring bright optical emission lines, and therefore will only be achieved with future observatories such as JWST and ELTs. Until then, the study of galaxies at lower redshifts, for which both the rest-frame UV and optical can be measured with existing facilities, is fundamental to test our models and gain insights into their physical properties. Along these lines, 

\section{Data and modelling approach}

\subsection{Spectro-photometric observations of Local `analogues' of high-redshift galaxies }

\begin{figure*}
	\captionsetup[subfigure]{width=0.9\textwidth}
	\centering
	\begin{subfigure}{.48\hsize}
		\resizebox{\hsize}{!}{\includegraphics{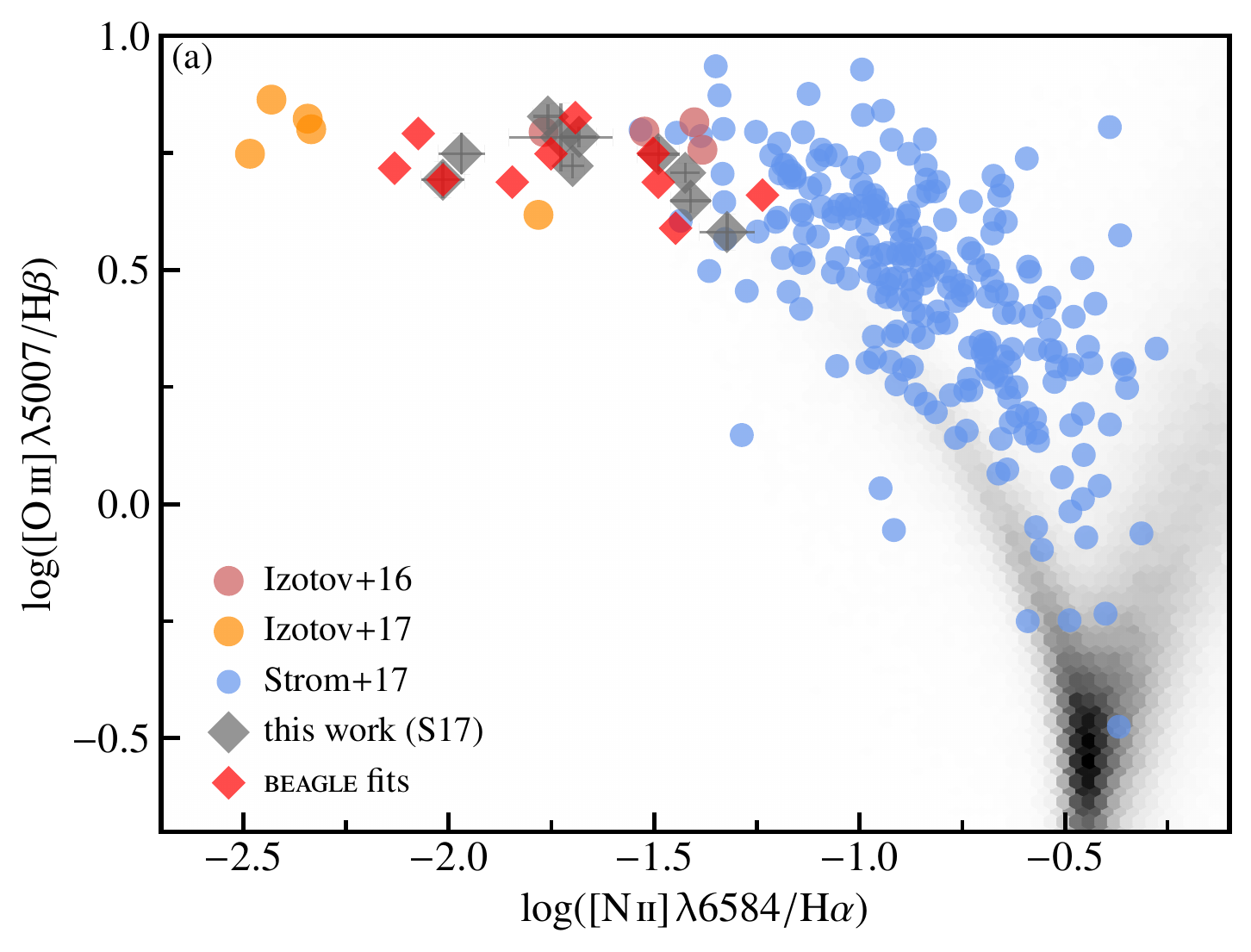}}
	\end{subfigure}
	\begin{subfigure}{.48\hsize}
	\resizebox{\hsize}{!}{\includegraphics{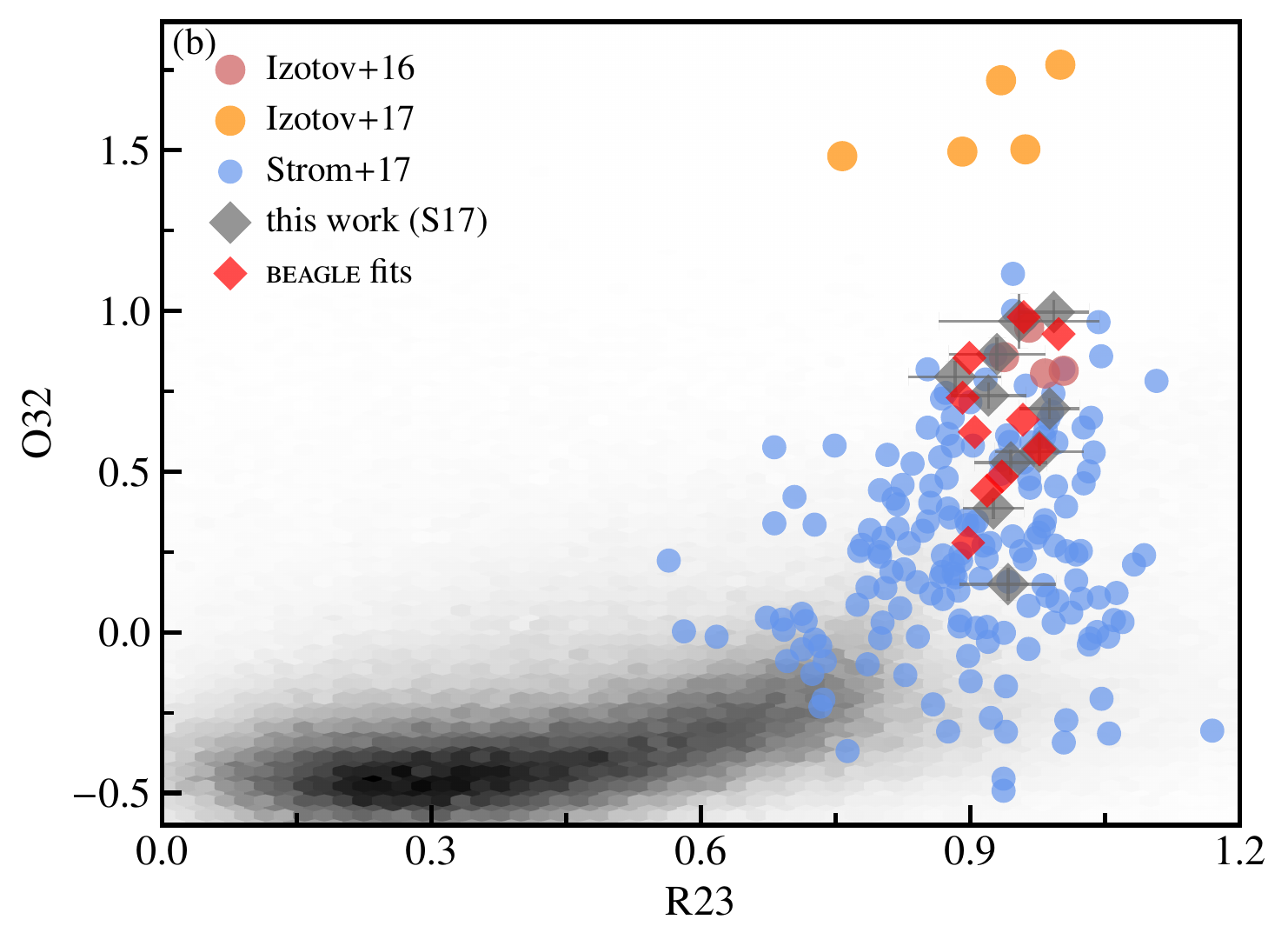}}
	\end{subfigure}	
	\caption{(a) \NIInoL\ BPT diagram showing star-forming galaxies at \range{z}{2}{2.6} from the KBSS survey (blue circles, \citealt{Strom2017}), the 4 LyC continuum `leaker' at $z\sim0.3$ from \citet{Izotov2016} (red circles), the 5 compact star-forming galaxies at $z<0.07$, candidates LyC continuum leakers, from \citet{Izotov2017b}, and the ten galaxies presented in S17 analysed in this work (grey diamonds), along with the model predictions obtained with \beagle\ (red diamonds). As a reference, we also show with a grey-scale (linear) density plot galaxies from the MPA/JHU emission line catalogue based on SDSS DR7 data. MPA/JHU fluxes are corrected for Galactic extinction only, while the other fluxes for internal extinction as well. (b) same as (a), but showing the $\txn{O32}=\log\left( \OIII/\OII\right)$ vs $\txn{R23}=\log{\left(\OIII+\OII)/\Hb\right)}$ diagram.}
	\label{fig:diagnostics}
\end{figure*}

%In this work we analyse the sample of 10 nearby galaxies with HST/COS, MMT, Keck/ESI, and SDSS observations presented in \citet{Senchyna2017} (hereafter S17). 

We appeal to a sample of ten nearby analogs of primeval galaxies, for which our team has acquired high-quality UV data from \textit{HST}/COS (G160M and G185M gratings) and optical spectra from MMT (see S17 for details). These galaxies were selected from a sample of \HeIIopt\ emitters dominated by stellar photoionization, identified by \citet{Shirazi2012} from the SDSS DR7 \citep{Abazajian2009}. These objects show extremely large equivalent widths of optical emission lines (e.g. $\OIII \gtrsim1000$ \AA), which, although rare in the nearby Universe (only $\sim 0.01$ per cent of SDSS galaxies), they become far more common at higher redshifts \citep[e.g.][]{Smit2014, Rasappu2016}. 
The ten galaxies in our sample have hard ionizing spectra able to doubly ionize He, i.e. with substantial flux at $\lambda < 228$ \AA\ (54.4 eV), although only half show Wolf-Rayet (WR) spectral features. Our follow-up \textit{HST} observations of these objects succeeded in revealing high-ionization UV lines (\CIV, \CIII, \OIIIuv) similar to those typically observed in galaxies at $z\gtrsim 5$ \citep{Stark2015a, Stark2015b, Stark2017, Mainali2017, Schmidt2017, Smit2017}, providing a unique reference sample for high-redshift studies. 

The location of our galaxies in the `\NIInoL' \citet[BPT]{Baldwin1981} diagram shown in Fig.~\ref{fig:diagnostics}a is typical of sub-solar metallicity gas (\range{\Z/\Zsun}{0.1}{0.7}) with a high ionization parameter (see, e.g., fig.~2 of \citealt{Gutkin2016}). The large O32 values at fixed R23, shown in Fig.~\ref{fig:diagnostics}b, also indicate a high ionization parameter, although these values can also be caused by density-bounded \HII\ regions \citep[e.g.][]{Jaskot2013, Stasinska2015}. As we show below, ionization-bounded regions are favoured by our model, which well matches the observed \OI\ fluxes, which should appear weaker for density-bounded regions.
Fig.~\ref{fig:diagnostics}a also shows that our galaxies have lower $\NIIb/\Ha$, at fixed $\OIIIa/\Hb$, than those in the KBSS sample of \citet{Strom2017}. These ratios are similar to those of the $z\sim0.3$ LyC leakers of \citet{Izotov2016}, although they do not reach the extreme $\NIIb/\Ha\sim-2.5$ of the compact star-forming galaxies of \citet{Izotov2017b}. Interestingly, Fig.~\ref{fig:diagnostics}b shows no significant offset between our galaxies, the KBSS sample and the LyC leakers in the O32 vs R23 diagram, while \citet{Izotov2017b} specifically targeted extreme O32 emitters. We also note the presence of a broad (a few $\times\,10^2 \, \txn{km}/\txn{s}$) component in several emission lines (\OIII, \Hb, \Ha), at a level of a few per cent the intensity of the narrow component. Such low-intensity broad components have been observed in other low-mass, low-metallicity, highly star-forming galaxies and linked to energy injection from stellar winds and supernovae (e.g. \citealt{Izotov2007, James2009, Amorin2012}), since Active Galactic Nuclei (AGNs) typically produce broader components ($\gtrsim 10^3 \, \txn{km}/\txn{s}$) with intensities comparable to the narrow ones \citep[e.g.][]{Izotov2008, Izotov2010}. 

\subsection{Emission line fluxes measurements}\label{sec:line_measure}

\begin{table*}
	\centering
	\begin{tabular}{c l l l l l l l l l l}
\toprule

 & \multicolumn{10}{c}{$\log{[F/F(\Ha)]}$}   \\

\cmidrule(lr){2-11} 

\multicolumn{1}{r}{ID}   & \multicolumn{2}{c}{\#2}  & \multicolumn{2}{c}{\#36}  & \multicolumn{2}{c}{\#80}  & \multicolumn{2}{c}{\#82}  & \multicolumn{2}{c}{\#110}\\

%\multicolumn{1}{c}{ID}	    &  \multicolumn{1}{c}{\logOH}  & \multicolumn{1}{c}{\logUs} &  \multicolumn{1}{c}{$\log \xiionst$}  &  \multicolumn{1}{c}{$\log \xiion$}  \\     

\multicolumn{1}{c}{Line}   & \multicolumn{1}{c}{Data}  & \multicolumn{1}{c}{Model}  & \multicolumn{1}{c}{Data}  & \multicolumn{1}{c}{Model}  & \multicolumn{1}{c}{Data}  & \multicolumn{1}{c}{Model}  & \multicolumn{1}{c}{Data}  & \multicolumn{1}{c}{Model}   & \multicolumn{1}{c}{Data}  & \multicolumn{1}{c}{Model} \\

\cmidrule(lr){1-1}  \cmidrule(lr){2-3}  \cmidrule(lr){4-5}  \cmidrule(lr){6-7} \cmidrule(lr){8-9}  \cmidrule(lr){10-11} 

$\textnormal{H}\delta$ & $-1.17\pm0.07$ & $-1.18$ & $-1.10\pm0.03$ & $-1.09$ & $-1.14\pm0.06$ & $-1.12$ & $-1.12\pm0.04$ & $-1.11$ & $-1.08\pm0.04$ & $-1.10$ \\
 $\textnormal{H}\gamma$ & $-0.90\pm0.07$ & $-0.89$ & $-0.82\pm0.03$ & $-0.82$ & $-0.86\pm0.06$ & $-0.84$ & $-0.84\pm0.04$ & $-0.84$ & $-0.81\pm0.04$ & $-0.82$ \\
 $\textnormal{H}\beta$ & $-0.54\pm0.08$ & $-0.54$ & $-0.48\pm0.03$ & $-0.48$ & $-0.52\pm0.07$ & $-0.50$ & $-0.49\pm0.04$ & $-0.50$ & $-0.48\pm0.03$ & $-0.47$ \\
 $\textnormal{H}\alpha$ & $0.00\pm0.1$ & $0.00$ & $0.00\pm0.04$ & $0.00$ & $0.00\pm0.08$ & $0.00$ & $0.00\pm0.06$ & $0.00$ & $0.00\pm0.05$ & $0.00$ \\
 $\textnormal{O}\,\textsc{iii}]\,\uplambda{1661}$ & $-2.14\pm0.07$ & $-2.21$ & $-2.02\pm0.04$ & $-1.98$ & $-2.51\pm0.07$ & $-2.47$ & $-1.81\pm0.04$ & $-1.90$ & $$ & $-2.08$ \\
 $\textnormal{O}\,\textsc{iii}]\,\uplambda{1666}$ & $-1.76\pm0.07$ & $-1.78$ & $-1.53\pm0.03$ & $-1.55$ & $-2.06\pm0.06$ & $-2.03$ & $-1.47\pm0.04$ & $-1.47$ & $-1.64\pm0.04$ & $-1.65$ \\
 $\textnormal{C}\,\textsc{iii}]\,\uplambda{1908}$ & $-1.24\pm0.08$ & $-1.27$ & $-1.0\pm0.06$ & $-1.04$ & $$ & $-1.69$ & $-0.84\pm0.04$ & $-0.86$ & $$ & $-1.37$ \\
 $[\textnormal{Ne}\,\textsc{iii}]\,\uplambda{3868}$ & $-0.95\pm0.07$ & $-0.97$ & $-0.85\pm0.03$ & $-0.85$ & $-0.90\pm0.06$ & $-0.88$ & $-0.79\pm0.04$ & $-0.78$ & $-0.80\pm0.04$ & $-0.80$ \\
 $[\textnormal{O}\,\textsc{ii}]\,\uplambda3727$ & $-0.63\pm0.08$ & $-0.64$ & $-0.40\pm0.05$ & $-0.32$ & $-0.29\pm0.07$ & $-0.24$ & $-0.60\pm0.06$ & $-0.60$ & $-0.21\pm0.05$ & $-0.15$ \\
 $\textnormal{He}\,\textsc{i}\,\uplambda{3889}$ & $-1.66\pm0.07$ & $-1.72$ & $-1.56\pm0.04$ & $-1.60$ & $-1.62\pm0.06$ & $-1.65$ & $-1.67\pm0.04$ & $-1.65$ & $-1.56\pm0.05$ & $-1.65$ \\
 $[\textnormal{O}\,\textsc{iii}]\,\uplambda{4363}$ & $-1.48\pm0.07$ & $-1.57$ & $-1.49\pm0.04$ & $-1.60$ & $-1.80\pm0.07$ & $-1.82$ & $-1.37\pm0.04$ & $-1.47$ & $-1.61\pm0.05$ & $-1.71$ \\
 $[\textnormal{Ar}\,\textsc{iv}]\,\uplambda{4740}$ & $-2.54\pm0.1$ & $-2.77$ & $-2.68\pm0.3$ & $-2.89$ & $-3.06\pm0.3$ & $-2.94$ & $-2.32\pm0.09$ & $-2.55$ & $-2.62\pm0.3$ &
$-2.87$ \\
 $[\textnormal{O}\,\textsc{iii}]\,\uplambda{4959}$ & $-0.26\pm0.08$ & $-0.30$ & $-0.23\pm0.03$ & $-0.27$ & $-0.25\pm0.07$ & $-0.23$ & $-0.15\pm0.04$ & $-0.18$ & $-0.24\pm0.03$ & $-0.26$ \\
 $[\textnormal{O}\,\textsc{iii}]\,\uplambda{5007}$ & $0.21\pm0.1$ & $0.18$ & $0.25\pm0.04$ & $0.21$ & $0.24\pm0.08$ & $0.25$ & $0.34\pm0.05$ & $0.30$ & $0.23\pm0.04$ & $0.21$ \\
 $\textnormal{He}\,\textsc{i}\,\uplambda{5875}$ & $-1.46\pm0.08$ & $-1.44$ & $-1.44\pm0.03$ & $-1.40$ & $-1.40\pm0.06$ & $-1.38$ & $-1.43\pm0.04$ & $-1.42$ & $-1.45\pm0.04$ & $-1.40$ \\
 $[\textnormal{O}\,\textsc{i}]\,\uplambda{6363}$ & $-2.69\pm0.1$ & $-2.63$ & $-2.40\pm0.2$ & $-2.41$ & $-2.47\pm0.09$ & $-2.35$ & $-2.63\pm0.09$ & $-2.71$ & $-2.52\pm0.1$ & $-2.44$ \\
 $[\textnormal{N}\,\textsc{ii}]\,\uplambda{6548}$ & $-2.52\pm0.1$ & $-2.61$ & $-2.49\pm0.3$ & $-2.32$ & $-2.00\pm0.08$ & $-1.97$ & $-2.38\pm0.2$ & $-2.55$ & $-1.93\pm0.07$ &
$-1.97$ \\
 $[\textnormal{N}\,\textsc{ii}]\,\uplambda{6583}$ & $-1.97\pm0.08$ & $-2.13$ & $-1.70\pm0.03$ & $-1.84$ & $-1.49\pm0.06$ & $-1.50$ & $-1.76\pm0.04$ & $-2.07$ & $-1.42\pm0.04$ & $-1.49$ \\
 $[\textnormal{S}\,\textsc{ii}]\,\uplambda{6716}$ & $-1.59\pm0.07$ & $-1.64$ & $-1.38\pm0.03$ & $-1.42$ & $-1.33\pm0.06$ & $-1.31$ & $-1.59\pm0.04$ & $-1.65$ & $-1.23\pm0.04$ & $-1.27$ \\
 $[\textnormal{S}\,\textsc{ii}]\,\uplambda{6730}$ & $-1.70\pm0.07$ & $-1.75$ & $-1.50\pm0.03$ & $-1.54$ & $-1.43\pm0.06$ & $-1.42$ & $-1.69\pm0.04$ & $-1.77$ & $-1.37\pm0.04$ & $-1.38$ \\
 $[\textnormal{Ar}\,\textsc{iii}]\,\uplambda{7136}$ & $-1.75\pm0.07$ & $-1.73$ & $-1.69\pm0.03$ & $-1.64$ & $-1.49\pm0.06$ & $-1.47$ & $-1.66\pm0.04$ & $-1.65$ & $-1.55\pm0.04$ & $-1.49$ \\

\bottomrule

\end{tabular}
\caption{UV and optical emission lines fluxes computed from the \HST/COS, MMT, and SDSS spectra, along with the fluxes predicted by our model. The model fluxes correspond to the posterior median of the marginal posterior distribution of each line flux. Although in our analysis we fitted the individual flux measurements, in this table, to facilitate the intepretation, we express the fluxes as $\log{[F/F(\Ha)]}$, where $F$ is the flux of any emission line and $F(\Ha)$ is the observed \Ha\ flux. The complete table is available online.}
\label{tab:lines_A}	
\end{table*}

\begin{table*}
	\centering
	\begin{tabular}{c l l l l l l l l l l}
\toprule

 & \multicolumn{10}{c}{$\log{[F/F(\Ha)]}$}   \\

\cmidrule(lr){2-11} 

\multicolumn{1}{r}{ID}   & \multicolumn{2}{c}{\#111} & \multicolumn{2}{c}{\#179} & \multicolumn{2}{c}{\#182} & \multicolumn{2}{c}{\#191} & \multicolumn{2}{c}{\#198}\\

%\multicolumn{1}{c}{ID}	    &  \multicolumn{1}{c}{\logOH}  & \multicolumn{1}{c}{\logUs} &  \multicolumn{1}{c}{$\log \xiionst$}  &  \multicolumn{1}{c}{$\log \xiion$}  \\     

\multicolumn{1}{c}{Line}   & \multicolumn{1}{c}{Data}  & \multicolumn{1}{c}{Model}  & \multicolumn{1}{c}{Data}  & \multicolumn{1}{c}{Model}  & \multicolumn{1}{c}{Data}  & \multicolumn{1}{c}{Model}  & \multicolumn{1}{c}{Data}  & \multicolumn{1}{c}{Model}   & \multicolumn{1}{c}{Data}  & \multicolumn{1}{c}{Model} \\

\cmidrule(lr){1-1}  \cmidrule(lr){2-3}  \cmidrule(lr){4-5}  \cmidrule(lr){6-7} \cmidrule(lr){8-9}  \cmidrule(lr){10-11} 

$\textnormal{H}\delta$ & $-1.10\pm0.04$ & $-1.06$ & $$ & $-1.24$ & $-1.11\pm0.04$ & $-1.08$ & $-1.06\pm0.1$ & $-1.06$ & $-1.09\pm0.07$ & $-1.08$ \\
 $\textnormal{H}\gamma$ & $-0.82\pm0.04$ & $-0.80$ & $-0.89\pm0.06$ & $-0.87$ & $-0.85\pm0.04$ & $-0.81$ & $-0.80\pm0.1$ & $-0.80$ & $-0.84\pm0.07$ & $-0.81$ \\
 $\textnormal{H}\beta$ & $-0.48\pm0.04$ & $-0.47$ & $-0.54\pm0.06$ & $-0.49$ & $-0.50\pm0.04$ & $-0.48$ & $-0.47\pm0.1$ & $-0.47$ & $-0.50\pm0.08$ & $-0.47$ \\
 $\textnormal{H}\alpha$ & $0.00\pm0.04$ & $0.00$ & $0.00\pm0.09$ & $0.00$ & $0.00\pm0.06$ & $0.00$ & $0.00\pm0.2$ & $0.00$ & $0.00\pm0.1$ & $0.00$ \\
 $\textnormal{O}\,\textsc{iii}]\,\uplambda{1661}$ & $-2.03\pm0.05$ & $-1.79$ & $$ & $-2.16$ & $-2.00\pm0.04$ & $-2.03$ & $-2.29\pm0.1$ & $-2.34$ & $$ & $-2.35$ \\
 $\textnormal{O}\,\textsc{iii}]\,\uplambda{1666}$ & $-1.35\pm0.03$ & $-1.36$ & $-1.93\pm0.06$ & $-1.92$ & $-1.65\pm0.04$ & $-1.60$ & $$ & $-1.91$ & $$ & $-1.90$ \\
 $\textnormal{C}\,\textsc{iii}]\,\uplambda{1908}$ & $$ & $-1.32$ & $-1.23\pm0.06$ & $-1.19$ & $-0.93\pm0.04$ & $-0.90$ & $-1.16\pm0.1$ & $-1.19$ & $-1.40\pm0.08$ & $-1.37$ \\
 $[\textnormal{Ne}\,\textsc{iii}]\,\uplambda{3868}$ & $-0.88\pm0.04$ & $-0.82$ & $-1.05\pm0.06$ & $-1.0$ & $-0.84\pm0.04$ & $-0.81$ & $-0.81\pm0.1$ & $-0.83$ & $-1.02\pm0.07$ & $-0.97$ \\
 $[\textnormal{O}\,\textsc{ii}]\,\uplambda3727$ & $-0.50\pm0.05$ & $-0.39$ & $-0.26\pm0.07$ & $-0.15$ & $-0.36\pm0.06$ & $-0.28$ & $-0.54\pm0.1$ & $-0.45$ & $-0.00.1\pm0.08$ & $-0.04$ \\
 $\textnormal{He}\,\textsc{i}\,\uplambda{3889}$ & $-1.60\pm0.05$ & $-1.58$ & $$ & $-1.90$ & $-1.65\pm0.04$ & $-1.60$ & $-1.49\pm0.1$ & $-1.64$ & $-1.53\pm0.08$ & $-1.62$ \\
 $[\textnormal{O}\,\textsc{iii}]\,\uplambda{4363}$ & $-1.43\pm0.05$ & $-1.52$ & $-2.02\pm0.07$ & $-2.09$ & $-1.49\pm0.04$ & $-1.61$ & $-1.80\pm0.1$ & $-1.83$ & $-1.98\pm0.09$ & $-1.89$ \\
 $[\textnormal{Ar}\,\textsc{iv}]\,\uplambda{4740}$ & $-2.55\pm0.3$ & $-2.79$ & $-3.59\pm1$ & $-3.09$ & $-2.72\pm0.2$ & $-2.83$ & $-2.78\pm0.2$ & $-2.66$ & $-3.51\pm1$ & $-3.28$ \\
 $[\textnormal{O}\,\textsc{iii}]\,\uplambda{4959}$ & $-0.26\pm0.04$ & $-0.25$ & $-0.35\pm0.06$ & $-0.31$ & $-0.19\pm0.04$ & $-0.21$ & $-0.15\pm0.1$ & $-0.12$ & $-0.38\pm0.08$ & $-0.36$ \\
 $[\textnormal{O}\,\textsc{iii}]\,\uplambda{5007}$ & $0.22\pm0.05$ & $0.22$ & $0.12\pm0.07$ & $0.17$ & $0.29\pm0.05$ & $0.27$ & $0.32\pm0.2$ & $0.36$ & $0.08\pm0.09$ & $0.12$ \\
 $\textnormal{He}\,\textsc{i}\,\uplambda{5875}$ & $-1.49\pm0.04$ & $-1.41$ & $-1.42\pm0.06$ & $-1.37$ & $-1.44\pm0.04$ & $-1.39$ & $-1.39\pm0.1$ & $-1.35$ & $-1.46\pm0.08$ & $-1.39$ \\
 $[\textnormal{O}\,\textsc{i}]\,\uplambda{6363}$ & $-2.75\pm0.3$ & $-2.52$ & $-2.61\pm0.1$ & $-2.45$ & $-2.41\pm0.08$ & $-2.39$ & $-2.90\pm0.2$ & $-2.68$ & $-2.45\pm0.1$ & $-2.24$ \\
 $[\textnormal{N}\,\textsc{ii}]\,\uplambda{6548}$ & $-2.53\pm0.2$ & $-2.49$ & $-1.95\pm0.09$ & $-1.73$ & $-2.22\pm0.07$ & $-2.22$ & $-2.23\pm0.1$ & $-2.16$ & $-1.85\pm0.09$ & $-1.93$ \\
 $[\textnormal{N}\,\textsc{ii}]\,\uplambda{6583}$ & $-2.01\pm0.06$ & $-2.01$ & $-1.41\pm0.07$ & $-1.24$ & $-1.68\pm0.05$ & $-1.75$ & $-1.73\pm0.1$ & $-1.69$ & $-1.32\pm0.08$ & $-1.45$ \\
 $[\textnormal{S}\,\textsc{ii}]\,\uplambda{6716}$ & $-1.51\pm0.03$ & $-1.53$ & $-1.29\pm0.06$ & $-1.26$ & $-1.37\pm0.04$ & $-1.41$ & $-1.62\pm0.1$ & $-1.62$ & $-1.21\pm0.07$ & $-1.19$ \\
 $[\textnormal{S}\,\textsc{ii}]\,\uplambda{6730}$ & $-1.68\pm0.04$ & $-1.64$ & $-1.41\pm0.06$ & $-1.37$ & $-1.48\pm0.04$ & $-1.52$ & $-1.76\pm0.1$ & $-1.73$ & $-1.35\pm0.07$ & $-1.30$ \\
 $[\textnormal{Ar}\,\textsc{iii}]\,\uplambda{7136}$ & $-1.76\pm0.04$ & $-1.70$ & $-1.50\pm0.06$ & $-1.47$ & $-1.73\pm0.04$ & $-1.58$ & $-1.43\pm0.1$ & $-1.51$ & $-1.56\pm0.08$ & $-1.56$ \\

\bottomrule

\end{tabular}
\caption{Continues from Table~\ref{tab:lines_A}.}
\label{tab:lines_B}	
\end{table*}

We measure the fluxes of the emission lines listed in Table~\ref{tab:lines_A}--\ref{tab:lines_B} and their associated errors with a custom fitting software described in S17. In summary, for each line, or group of neighbouring lines, we consider a linear function to describe the local continuum and fit the line profile with a Gaussian function. For lines with multiple kinematic components, we consider a second Gaussian function. We do not attempt to subtract the underlying stellar absorptions from the measured line fluxes, since stellar absorptions are self-consistently included in our stellar population + photoionization model (see Section~\ref{sec:modelling} below). We rescale the formal errors on the line fluxes measured from SDSS spectra by a factor 1.8, the average error scaling suggested by the MPA/JHU analysis.\footnote{\url{https://wwwmpa.mpa-garching.mpg.de/SDSS/DR7/raw_data.html}} As in S17, we correct the line fluxes for Galactic extinction using the dust map of \citet{Schlafly2011} and assuming the $R_V = 3.1$ extinction curve of \citet{Fitzpatrick1999}. Since the predictions of our model include only H-Balmer transitions of principal quantum number $n\le6$, where $n=6$ corresponds to \Hd, we subtract from the measured \HeILam{3889} flux the contribution from the \Hz\lam{3889} line using the measured \Hd\ flux, and assuming a case B recombination ratio of $\Hz/\Hd = 0.41$ valid for a $T = 10^4 \, \txn{K}$ gas \citep[see Table~4.2 of][]{Osterbrock2006}.\footnote{The proximity in wavelength of \Hd\ at $\uplambda = 4101.7$ \AA\ to \Hz\lam{3889} allows us to ignore the potential impact of dust attenuation on the estimation of the \Hz\ flux, since \Hd\ and \Hz\ would suffer a comparable amount of attenuation.} Given the similar (circular) apertures of SDSS fibers and HST/COS (diameters of $3 \arcsec$ and $2.5 \arcsec$, respectively), we do not apply any aperture correction to the UV and optical emission lines measurements.
 
\subsection{Stellar-population and photoionization modelling}\label{sec:modelling}

We model the UV-to-optical photometry and emission lines integrated fluxes of our galaxies to constrain their physical properties and production rate of H-ionizing photons. To achieve this, we use \beagle, a new-generation spectral analyses tool which incorporates a self-consistent description of the emission from stars (based on the latest version of the \citealt{Bruzual2003} stellar population synthesis code\footnote{See appendix~A of \citealt{Vidal2017} for a comparison of the population synthesis code used in this work with the original version of \citet{Bruzual2003}.}) and photoionized gas (using \cloudy\ version 13.3, last described in \citealt{Ferland2013}). 
In practice, we consider a \citet{Chabrier2003} stellar initial mass function truncated at 0.1 and 100 \Msun, and adopt a constant star formation rate and a constant metallicity. The star formation and chemical enrichment history of our galaxies is hence parametrised in terms of the galaxy stellar mass \M, age of the oldest stars \t\ (or age of the galaxy) and stellar metallicity \Z. Following S17, we account for the effect of dust attenuation by considering the Small Magellanic Cloud extinction curve \citep{Pei1992}, and vary the $V$-band attenuation optical depth \tauV. The choice of using an extinction curve is justified by the fact that geometry and scattering effects should have only a minor influence on line emission from the small physical regions probed by our data.

\begin{table}
	\centering
	\begin{tabular}{C{0.30\columnwidth-2\tabcolsep} C{0.3\columnwidth-2\tabcolsep} L{0.4\columnwidth-2\tabcolsep}}
\toprule

%\multirow{2}{*}{Parameter}	    &  \multicolumn{2}{c}{Prior}  &  \multicolumn{1}{c}{\multirow{2}{*}{Description}} \\     

\multicolumn{1}{c}{Parameter}	    &  \multicolumn{1}{c}{Prior} &  \multicolumn{1}{c}{Description}  \\     

\midrule

\logM		          & 	 $\mathcal{U} \in [1,9]$  & Stellar mass  \\

\logt 		    & 		$\mathcal{N}(7.5;\,1.5)$  $\in [7, 10.15]$ &    Age of oldest stars in the galaxy  \\

\logZ		        & 	 $\mathcal{U}\in [-2.2,0.25]$ &  Stellar and interstellar metallicity   \\

\logUs		                    & 	 $\mathcal{U}\in [-4,-1]$ &  	 Effective gas ionization parameter \\

$\tauV$		                    & 	 $\exp{-\tauV}$  $\in [0, 2]$ &  	 $V$-band attenuation optical depth \\

%\logtburst 		    & 		$\mathcal{N}(6.5;\,0.5)$  $\in [6, 7]$ &    Duration of the most recent burst of star formation  \\

%\logsfr		               & 		$\mathcal{N}(-1;\,2)$  $\in [-4, 4]$ &   Star formation rate during the most recent burst  \\

%\logZism		        & 	 $\mathcal{U}\in [-2.2,0.25]$ &  Interstellar metallicity and metallicity of stars with ages $\tprime < 10^7$ yr  \\

\CO		                    &  $\mathcal{U}\in [0.1,1]$ & 	  Carbon-to-oxygen ratio \\

$\xid$		                    &  $\mathcal{U}\in [0.1,0.5]$ & 	  Dust-to-metal mass ratio \\

\bottomrule
	\end{tabular}
\caption{Free parameters of the model adopted in \beagle\ to fit the COS/SDSS photometry + Balmer emission lines [parameters \logM, \logt, \logZ, \logUs\ and \tauV], and to fit the emission line fluxes (all parameters in the table) of the ten nearby star-forming galaxies analysed in this work. The symbol $\mathcal{U}$ indicates a uniform distribution in the specified range, $\mathcal{N}(\mu;\,\sigma)$ a Gaussian (Normal) distribution with mean $\mu$ and standard deviation $\sigma$, truncated in the specified range.}
\label{tab:priors}	
\end{table}

We follow \citet{Gutkin2016} to account for (line+continuum) nebular emission. We fix the gas density at $\nH = 10^2 \, \txn{cm}^{-3}$, the typical value observed in local compact star-forming galaxies \citep[e.g.][]{Izotov2006, Jaskot2013}, set the interstellar metallicity equal to the stellar metallicity ($\Zism = \Z$), and vary the ionization parameter \logUs, which sets the ratio of H-ionizing-photon to gas density at the Str\"omgren radius,\footnote{This parametrization translates into a volume-averaged ionization parameter $\langle U \rangle=9/4\,\Us$ \citep[see equation 1 of][]{Hirschmann2017}.} the dust-to-metal mass ratio \xid, which sets the depletion of heavy elements onto dust grains, and the carbon-to-oxygen abundance ratio \CO, where $\CO = 1$ implies a scaled-Solar abundance.

We perform two separate fittings with \beagle: one to a combination of UV and optical photometry and Balmer emission lines, to derive the intrinsic UV luminosity of our galaxies; the other to a broad set of UV and optical emission lines, to derive the physical properties of the ionized gas and ionization sources. In the combined photometry+Balmer lines fit, we consider the H-Balmer lines \Ha, \Hb, \Hg\ and \Hd, the $ugri$ optical (fiber) photometry from SDSS,\footnote{We do not fit the reddest SDSS band ($z$) since the adopted (constant) star formation histories may introduce biases when fitting a band  potentially dominated by evolved stellar populations.} and a single UV photometric band, which we compute from an integration of the \textit{HST}/COS spectra.\footnote{Note that we do not use GALEX photometry, to avoid potential biases introduced by the large GALEX PSF.} Because of the different filter configurations used when acquiring the COS data, we consider a flat filter covering the region \range{\lambda}{1620}{1660}\,\AA\ for the objects \SBID{36} and \SBID{182}, and a filter covering the range \range{\lambda}{1480}{1515}\,\AA\ for the other 8 objects. Both regions sample the rest-frame UV continuum of the galaxies, and are not significantly affected by nebular emission lines or ISM absorption lines. Since Balmer lines alone cannot constrain the several nebular emission parameters, we simplify the model and fix the carbon-to-oxygen abundance ratio to the Solar value [$\CO = (\CO)_\odot$] and the depletion factor to $\xid=0.3$, to obtain a model described by the 5 free parameters $\thetab^\txn{phot+Balmer} = [\logM, \logt, \logZ, \logUs, \tauV]$. 

In a separate fitting, we model the broad range of UV and optical emission lines listed in Table~\ref{tab:lines_A}--\ref{tab:lines_B}. Besides the H-Balmer lines \Ha, \Hb, \Hg\ and \Hd, we include several O, $\txn{O}^+$ and $\txn{O}^{2+}$ lines, the nitrogen and sulphur doublets \NII\ and \SII, the \CIII\ line, helium lines, and lines from $\txn{Ar}^{2+}$, $\txn{Ar}^{3+}$ and $\txn{Ne}^{2+}$. We fit with \beagle\ the integrated line fluxes, computed from the observed spectra as detailed in Section~\ref{sec:line_measure}, with a model similar to the one described above, but letting the carbon-to-oxygen abundance ratio and depletion factor free to vary. This model is therefore described by the 7 free parameters $\thetab^\txn{lines} = [\logM, \logt, \logZ, \logUs, \tauV, \CO, \xid]$.

The Bayesian approach adopted in \beagle\ allows us to derive the posterior probability distribution of the model parameters $\thetab^\txn{phot+Balmer}$ and $\thetab^\txn{lines}$, 
\begin{equation}\label{eq:posterior}
\conditional{\thetab}{\Db,H} \propto \prior\ \, \likelihood \, ,
\end{equation}
where \Db\ indicates the data, $H$ the adopted model, \prior\ the prior distribution, and \likelihood\ the likelihood function. We adopt independent priors for each parameter, uniform for \logM, \logZ, \logUs, \xid, \CO, Gaussian for \logt, and exponential for \tauV, and report the prior definitions and ranges in Table~\ref{tab:priors}. We adopt a Gaussian likelihood function with independent errors on each measurement,
\begin{equation}\label{eq:likelihood}
- 2\, \ln \mathcal{L}(\thetab) = \sum_i \left [ \frac{y_i-\hat{y}_i(\thetab)}{\sigma_i} \right ]^2 \, ,    
\end{equation}
where the summation index $i$ runs over all observables considered in the fitting, i.e. the Balmer lines and COS+SDSS photometry in one case, the emission lines listed in Table~\ref{tab:lines_A}--\ref{tab:lines_B} in the other case. In equation~\ref{eq:likelihood}, $y_i$ indicates the data,  $\sigma_i$ the observational error and $\hat{y_i}(\bmath{\thetab})$ the model prediction. We adopt the Nested Sampling algorithm \citep{Skilling2006} in the \multinest\ implementation \citep{Feroz2009} to sample the posterior probability distribution of the free parameters of our model. We note that this implies that each model spectrum is computed in \beagle\ `on-the-fly' for each combination of parameters selected by \multinest\ during the sampling of the posterior probability distribution of the model parameters.

\section{Results}

\begin{figure*}
	\begin{subfigure}{.99\hsize}
	\resizebox{\hsize}{!}{\includegraphics{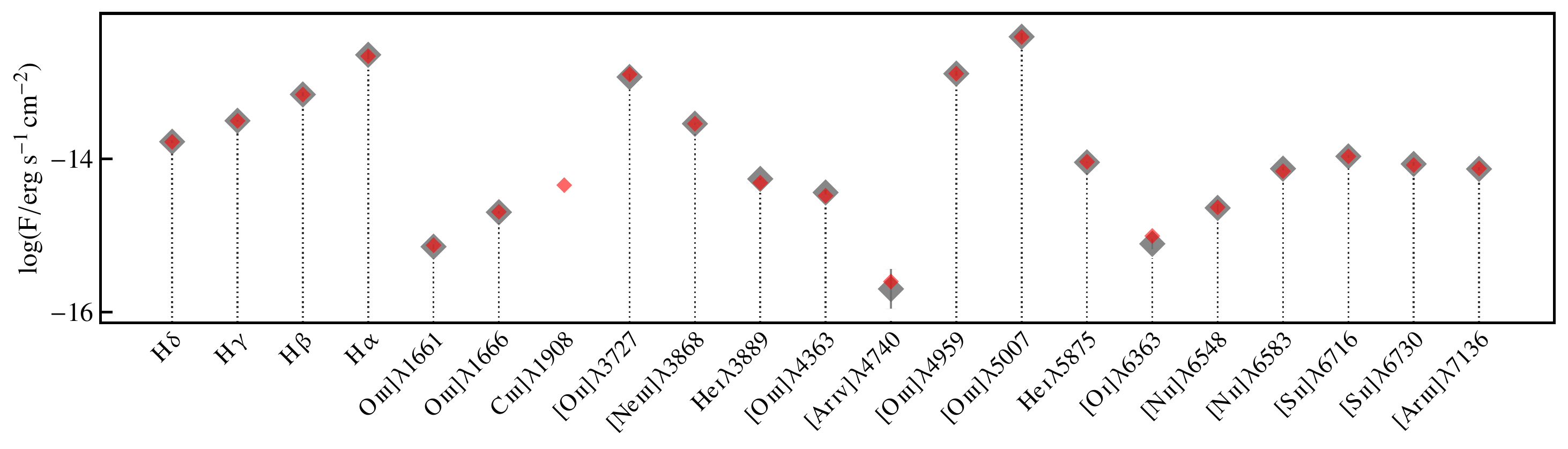}}
	\end{subfigure}		
	\caption{Example of a \beagle\ fitting of the galaxy \SBID{80}, an embedded \HII\ region. Observed fluxes are indicated by grey diamonds, model fluxes by red ones. Note that in most cases the error bars on the observed fluxes are contained within the markers.}
	\label{fig:fitting}
\end{figure*} 

The model we have adopted allows us to well reproduce all the observables here considered , i.e. the combination of COS/SDSS photometry and H-Balmer lines, and the emission lines of Table~\ref{tab:lines_A}--\ref{tab:lines_B}. In Table~\ref{tab:lines_A}--\ref{tab:lines_B}, we also report the emission line fluxes predicted by our model for each object fitted, while we show in Fig.~\ref{fig:fitting} a visual example of the fitting of the object \SBID{80}, an embedded \HII\ region at $z=0.01085$. We show in Figs~\ref{fig:diagnostics}a,b, the line ratios predicted by \beagle\ for all fitted objects in our sample. 
Table~\ref{tab:lines_A}--\ref{tab:lines_B} indicates a remarkable agreement between the predictions of our relatively simple model and the data.

\subsection{Physical properties of extreme star-forming regions}\label{sec:phys_properties}

\begin{table*}
	\centering
	\begin{tabular}{C{0.15\columnwidth-2\tabcolsep} C{0.25\columnwidth-2\tabcolsep}  C{0.25\columnwidth-2\tabcolsep} C{0.25\columnwidth-2\tabcolsep} C{0.25\columnwidth-2\tabcolsep}  C{0.25\columnwidth-2\tabcolsep}}
\toprule

\multicolumn{1}{c}{ID} & \multicolumn{1}{c}{$\t/\txn{Myr}$}  &  \multicolumn{1}{c}{\logOH}  & \multicolumn{1}{c}{\logUs} &  \multicolumn{1}{c}{\xid} &  \multicolumn{1}{c}{\tauV}  \\     

\midrule

 2 & $1.10\pm0.1$ & $7.87\pm0.02$ & $-2.56\pm0.02$ & $0.15\pm0.03$ & $0.37\pm0.01$ \\
 36 & $1.11\pm0.1$ & $7.97\pm0.03$ & $-2.77\pm0.02$ & $0.19\pm0.04$ & $0.1\pm0.02$ \\
 80 & $3.10\pm0.8$ & $8.31\pm0.03$ & $-2.65\pm0.03$ & $0.18\pm0.03$ & $0.17\pm0.03$ \\
 82 & $2.03\pm0.1$ & $7.95\pm0.01$ & $-2.36\pm0.02$ & $0.28\pm0.01$ & $0.18\pm0.008$ \\
 110 & $11.06\pm4$ & $8.15\pm0.03$ & $-2.66\pm0.02$ & $0.35\pm0.03$ & $0.04\pm0.02$ \\
 111 & $1.27\pm0.3$ & $7.90\pm0.03$ & $-2.68\pm0.02$ & $0.18\pm0.04$ & $0.04\pm0.02$ \\
 179 & $177.99\pm5$ & $8.49\pm0.02$ & $-2.60\pm0.03$ & $0.18\pm0.03$ & $0.01\pm0.01$ \\
 182 & $1.07\pm0.08$ & $8.10\pm0.03$ & $-2.68\pm0.02$ & $0.16\pm0.03$ & $0.08\pm0.02$ \\
 191 & $3.15\pm0.2$ & $8.47\pm0.01$ & $-2.14\pm0.03$ & $0.12\pm0.02$ & $0.00\pm0.005$ \\
 198 & $7.05\pm2$ & $8.20\pm0.08$ & $-2.95\pm0.06$ & $0.12\pm0.02$ & $0.02\pm0.02$ \\

\bottomrule
	\end{tabular}
\caption{Physical parameters of the ten objects analysed in this work, obtained by fitting with \beagle\ the emission lines listed in Table~\ref{tab:lines_A}--\ref{tab:lines_B}. We report the posterior median and 68 per cent central credible interval of stellar age \t, gas-phase metallicity \logOH, ionization parameter \logUs, dust-to-metal mass ratio \xid\ and $V$-band attenuation optical depth \tauV.}
\label{tab:physical_properties}	
\end{table*}

Table~\ref{tab:physical_properties} lists the physical properties of our sample derived from the \beagle-based emission lines fitting. As already shown in S17, our ten galaxies cover a broad range of gas-phase metallicities \range{\logOH}{7.9}{8.5} [\rangeto{\logZ}{-0.8}{-0.2}] and show low dust attenuation values. We derive high ionization parameters in all galaxies [$\logUs \gtrsim -3$], even in the highest metallicity ones, namely \SBID{179} and \SBID{191} [$\logOH\sim 8.5$], which show $\logUs \sim -2.60$ and $\sim -2.14$, respectively. Galaxies showing intermediate metallicities and high ionization parameters are rare in the Local Universe (e.g. among SDSS galaxies, see fig~2 of \citealt{Carton2017}), while they become more common at higher redshifts, as suggested by the detection of high-ionization metal lines at $z\gtrsim 5$ \citep[e.g.][]{Stark2015a, Mainali2017}. The emission lines that we consider in this analysis originate from both refractory (O, C) and non-refractory (N, S, Ar, Ne) elements, and this enables us to place tight constraints on the fraction of metals depleted onto dust grains \xid. The \xid\ values reported in Table~\ref{tab:physical_properties} are significantly lower than the Solar value $\xid_\odot = 0.36$, indicating that most metals in these galaxies are in the gas phase and not locked into dust grains.  

\begin{figure}
	\centering
	\resizebox{\hsize}{!}{\includegraphics{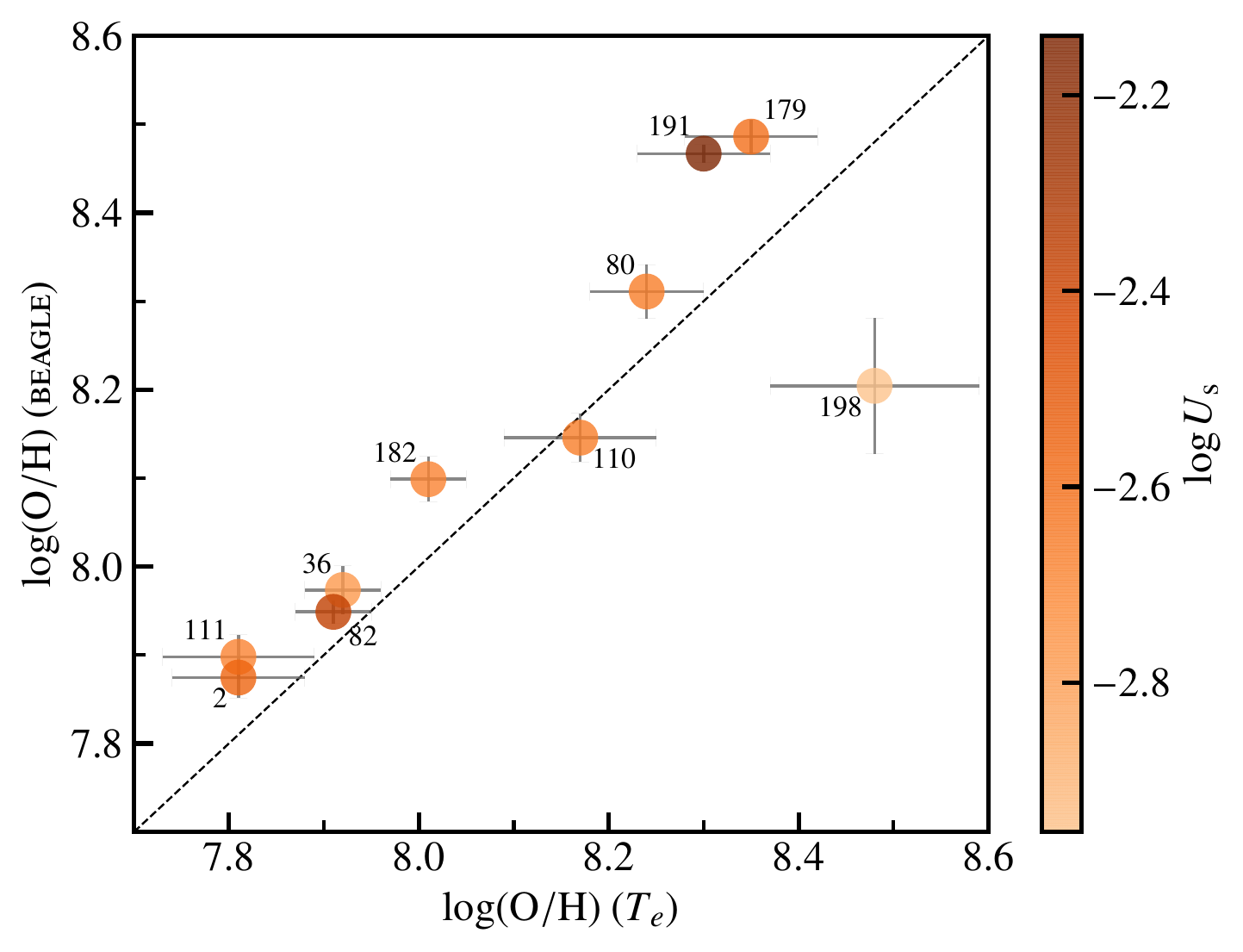}}
	\caption{Comparison between the gas-phase metallicity \logOH\ obtained in this work with the \beagle\ tool (y-axis) and by S17 with the `direct-\Te' method. Colour-coded is the ionization parameter \logUs\ obtained with the \beagle\ analysis. Error bars indicate the 68 per cent central credible interval.}
	\label{fig:logOH}
\end{figure}

Fig.~\ref{fig:logOH} compares the gas-phase metallicity \logOH\ derived with our analysis with the values obtained by S17 using a standard method, the `direct-\Te' one (see their section~3.2). S17 measure the electron temperature (\Te) of $\txn{O}^+$ and $\txn{O}^{2+}$ using ratios of auroral ([O\,{\sc ii]}\lamlam{7320}{7330} and [O\,{\sc iii]}\lam{4363}) and strong (\OII\ and \OIII) lines, and then compute the abundances of the $\txn{O}^+$ and $\txn{O}^{2+}$ ions using the strong lines. As Fig.~\ref{fig:logOH} shows, the agreement between the S17, direct-\Te\ based estimates of \logOH\ and our values (listed in Table~\ref{tab:physical_properties}) is generally good, although the \beagle-based metallicities are slightly larger than the S17 ones, by $\sim 0.05$ dex at $\logOH \lesssim 8.3$, and by $\sim0.2$ dex for the two galaxies with larger \logOH\ values, \SBID{179} and \SBID{191}. These two objects show large ionization parameters, and S17 ignore the contribution from $\txn{O}^{3+}$ when measuring the oxygen abundance, but they evaluate the contribution of $\txn{O}^{3+}$ to be minor by considering a relation between $\txn{O}^{3+}$ and $\txn{He}^{2+}/\txn{He}^{+}$ \citep{Izotov2006}. The only object that does not follow the trend above is \SBID{198}, for which our metallicity estimate is $\sim 0.3$ dex lower than the S17 one. The metallicity estimated for \SBID{198}, however, is the most uncertain, and the statistical significance of the difference among the two methods is $\sim 3 \, \sigma$, hence more data would be necessary to gain better insight into the origin of this discrepancy.

We note that differences between metallicities estimated with photoionization models and with the `direct-\Te' method, such as those discussed above, have been reported in several recent works \citep[e.g.][]{Blanc2015, ValeAsari2016, PerezMontero2017}. Interestingly, both \citet{Blanc2015} and \citet{ValeAsari2016} find that metallicities estimated with photoionization models are systematically larger (by $\sim 0.2$ to $0.4$ dex) than those estimated with the direct-\Te\ method, qualitatively consistent with our findings, and that  photoionization models can provide metallicity estimates in better agreement with those derived from metal recombination lines. In \citet{ValeAsari2016} (see their sec.~6.3) it is suggested that the [O\,{\sc iii]}\lam{4363} line is at the origin of the measured differences among photoionization- and direct-\Te- based metallicities. The [O\,{\sc iii]}\lam{4363} line is used to determine the electron temperature in the $\txn{O}^{2+}$ zone, but this line can be affected by temperature fluctuations and non-Maxwellian velocity distributions of the free electrons \citep{Nicholls2012}. These effects can boost the [O\,{\sc iii]}\lam{4363} line flux, thus modifying the derived electron temperature and, consequently, the metallicity. In this work, the differences among the photoionization- and direct-\Te- based metallicities are, for most objects, significantly smaller than found in previous works. The reason is likely related to the fact that S17 directly measure the \Te\ of both $\txn{O}^+$ and $\txn{O}^{2+}$, instead of only measuring the temperature of $\txn{O}^{2+}$ and using calibrated relations to derive that of $\txn{O}^+$. The S17 approach can thus alleviate some of the limitations of the direct-\Te\ method, providing estimates in better agreement with those derived with the photoionization models incorporated into \beagle.

%This may be an indication of the limitation of the `direct-\Te' method \citep[see also section 5.2 of][]{Gutkin2016} for such extreme objects, but more data points are necessary to perform a statistical analysis. 
%The self-consistent modelling of the UV and optical spectra of our objects, including stellar winds features, will be the subject of a follow-up study.

\subsection{Production rate of H-ionizing photons}\label{sec:xi_ion}

In this paper, we define $\xiionst = \Nion/\Luvst$ as the \emph{ratio between the production rate of H-ionizing photons (\Nion) and the intrinsic stellar monochromatic UV luminosity (\Luvst)}. The quantity \Luvst\ is computed from the pure stellar spectrum considering a flat filter centred at 1500 \AA\ and of 100 \AA\ width \citep[e.g. see][]{Robertson2013}. \Luvst\ is the emission from stars that would be observed in the absence of gas and dust in galaxies, i.e. it does not include the effect of absorption and emission from neutral and photoionized gas, nor of dust attenuation. Previous studies have adopted different definitions of the monochromatic UV luminosity, for example, by considering the stellar+nebular luminosity \LuvHII, corrected for dust attenuation outside \HII\ regions but not for dust attenuation inside \HII\ regions nor for recombination-continuum emission, or the observed luminosity \Luv. For the sake of clarity, we introduce here the symbols \xiionHII\ and \xiion\ to differentiate between our definition of \xiionst\ and the ratios $\xiionHII = \Nion/\LuvHII$ and $\xiion = \Nion/\Luv$.

\begin{table}
	\centering
	\begin{tabular}{C{0.15\columnwidth-2\tabcolsep} C{0.25\columnwidth-2\tabcolsep}  C{0.25\columnwidth-2\tabcolsep} C{0.25\columnwidth-2\tabcolsep} }
\toprule

\multicolumn{1}{c}{ID}	   &  \multicolumn{1}{c}{$\log \xiionst$}  &  \multicolumn{1}{c}{$\log \xiionHII$} &  \multicolumn{1}{c}{$\log \xiion$} \\     

\midrule

2          &  $25.84\pm0.05$ &  $25.81\pm0.11$ &  $26.42\pm0.10$ \\
36         &  $25.37\pm0.02$ &  $25.36\pm0.03$ &  $25.38\pm0.04$ \\
80         &  $25.73\pm0.01$ &  $25.69\pm0.01$ &  $26.26\pm0.03$ \\
82         &  $25.80\pm0.01$ &  $25.84\pm0.02$ &  $25.85\pm0.02$ \\
110        &  $25.24\pm0.02$ &  $25.26\pm0.03$ &  $25.27\pm0.04$ \\
111        &  $25.38\pm0.03$ &  $25.37\pm0.03$ &  $25.47\pm0.04$ \\
179        &  $25.50\pm0.03$ &  $25.52\pm0.04$ &  $25.84\pm0.03$ \\
182        &  $25.57\pm0.02$ &  $25.61\pm0.03$ &  $25.73\pm0.05$ \\
191        &  $25.82\pm0.02$ &  $25.94\pm0.04$ &  $26.06\pm0.03$ \\
198        &  $25.65\pm0.02$ &  $25.60\pm0.03$ &  $25.63\pm0.03$ \\

\bottomrule
	\end{tabular}
\caption{Production rate of H-ionizing photons per unit monochromatic UV luminosity [expressed in units of $\log(\txn{erg}^{-1}\,\txn{Hz})$]. We report different definitions of this quantity, using the intrinsic stellar luminosity (\xiionst), the stellar+nebular luminosity (\xiionHII), and the observed (i.e. attenuated) luminosity (\xiion, see Section~\ref{sec:xi_ion}). The reported values correspond to the posterior median, while the errors indicate to the 68 per cent central credible interval.}
\label{tab:results}	
\end{table}

To compute \xiionst, we combine the monochromatic UV luminosity (\Luvst) inferred from the fitting to COS/SDSS photometry + Balmer lines with the rate of H-ionizing photons (\Nion) obtained by fitting the different UV and optical emission lines. %Table~\ref{tab:results} summarises the main results of our analysis, the constraints on the gas-phase metallicity, ionization parameter, and production rate of H-ionizing photons per unit intrinsic UV luminosity \xiionst. We also report the rate of ionizing photons per unit `observed' UV luminosity $\xiion = \Nion/\Luv$, which does not incorporate corrections for the effects of gas and dust within, and of dust outside, photoionized regions. 
We report in Table~\ref{tab:results} the \xiionst\ values for the ten galaxies in our sample. Our galaxies exhibit $\log (\xiionst/\txn{erg}^{-1}\,\txn{Hz})$ values in the range $\sim25.2$ to $\sim 25.8$, i.e. generally larger then the values adopted in standard reionization models (25.2 to 25.4). We also report in Table~\ref{tab:results} the values we derive for \xiionHII\ and \xiion, to highlight the impact that different definitions of this quantity can have on the quoted values. In particular, \xiionHII\ can be either smaller or larger than \xiionst, depending on the relative strengths of dust attenuation inside  \HII\ regions (which lowers \LuvHII\ and raises \xiionHII; see e.g. object \SBID{191}) and recombination-continuum emission (which boosts \LuvHII\ and hence lowers \xiionHII; see e.g. \SBID{2}). Considering the observed (i.e. attenuated) UV luminosity \Luv\ generally implies larger \xiion\ values, but the difference between \xiionst\ and \xiion\ depends on the amount of dust attenuation affecting stellar and nebular emission outside \HII\ regions.  

\begin{figure}
	\centering
	\resizebox{\hsize}{!}{\includegraphics{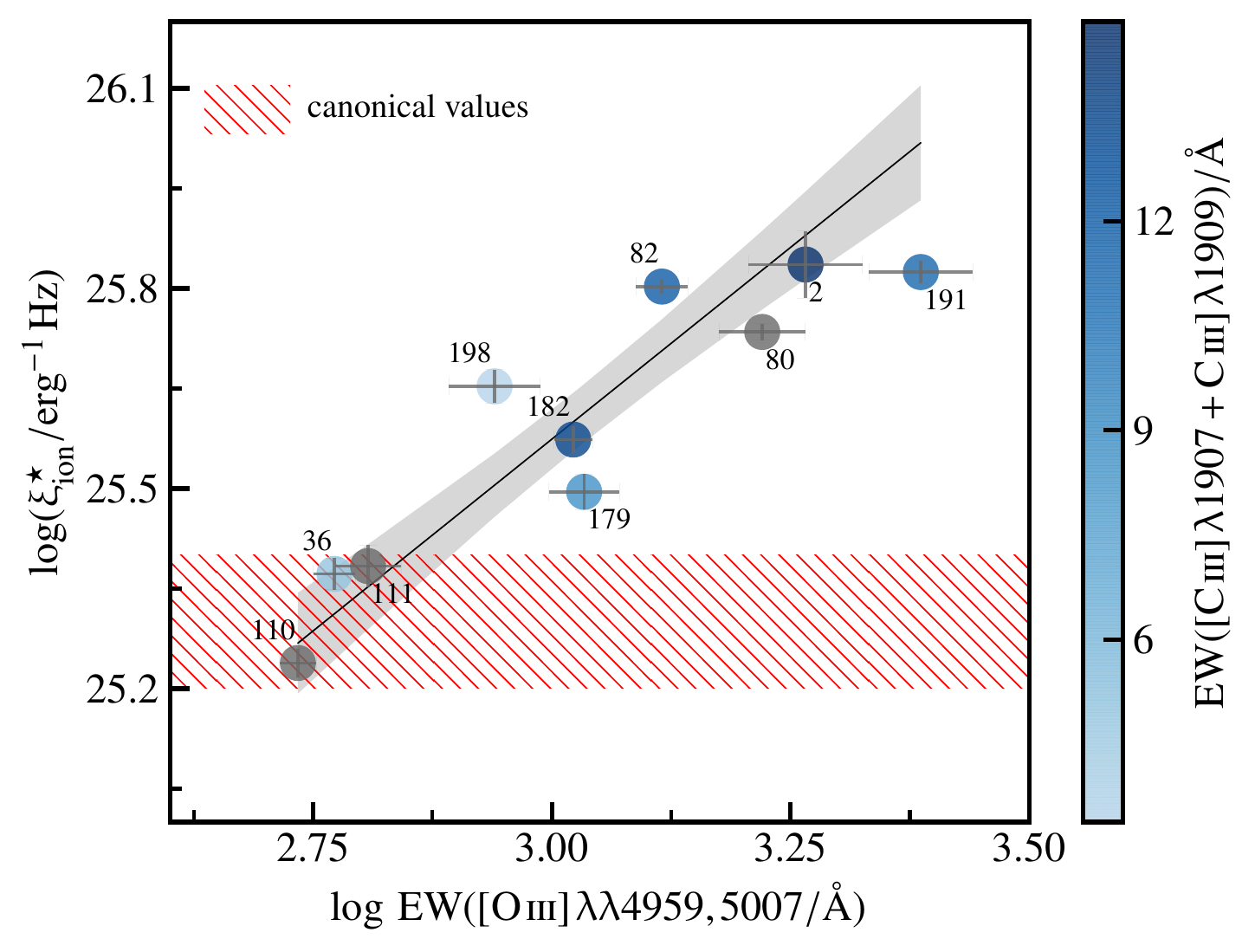}}
	\caption{Relation between production rate of H-ionizing photons per unit intrinsic monochromatic UV luminosity (y-axis) and equivalent width of the  \OIII\ doublet (x-axis). Colour-coded is the equivalent width of \CIII, while grey circles indicate objects with no \CIII\ detections. Error-bars on the circles indicate the 68 per cent central credible region. The black line indicates the best-fit linear relation between \xiionst\ and $\EW(\OIIInoL)$ reported in equation~\eqref{eq:xi_OIII}, while the grey bands indicate the 68 per cent credible region. The red hatched region indicates typical \xiionst\ values adopted in reionization models.}
	\label{fig:xi_ion}
\end{figure}

In Fig.~\ref{fig:xi_ion}, we show the relation between \xiionst\ and equivalent width of the \OIII\ doublet. Objects exhibiting larger $\EW(\OIIInoL)$ also show larger \xiionst. A relation between the two quantities is expected, since large \OIII\ equivalent widths are typically produced by stellar populations with young ages and sub-solar metallicities, which would also produce copious amounts of H-ionizing photons. Moreover, part of the small scatter visible in Fig.~\ref{fig:xi_ion} can be ascribed to the relatively narrow range of $\OIII/\Hb$ probed by our objects (Fig.~\ref{fig:diagnostics}a). Remarkably, the tight relation in Fig.~\ref{fig:xi_ion} also implies a limited variation of ionized-gas properties in our objects, since the relation between \xiionst\ and $\EW(\OIIInoL)$ depends on the shape of the ionizing spectrum (e.g. stellar age and metallicity), ionization parameter (geometry of the photoionized regions), metal abundances and gas density.
% small range of [O III]/Hbeta ratios in the sample (which are uniformly high). If so, I would suggest pointing this out as a caveat, noting that the [O III] relation should work reasonably well at high z for the sources with high [O III]/Hbeta 
The colour coding of the circles in Fig.~\ref{fig:xi_ion} further indicates the equivalent width of \CIII\ measured from the COS spectra. As noted by S17 (their fig.~10), the equivalent widths of \CIII\ and \OIII\ are related to each other, the most extreme \CIIInoL\ emitters showing largest \OIIInoL\ emission. Additional high-quality rest-frame UV spectra are required to calibrate a relation between \xiionst\ and \CIIInoL, but this relation has the potential to allow indirect constraints on \xiionst\ even for galaxies deep into the EoR.

We fit the relation between \xiionst\ and $\EW(\OIIInoL)$ using `Orthogonal Distance Regression', a linear regression method which allows us to account for errors along both the y- and x-axis.\footnote{\url{https://docs.scipy.org/doc/scipy/reference/odr.html}} The black line in Fig.~\ref{fig:xi_ion} show the best-fitting relation
\begin{equation}\label{eq:xi_OIII}
\log(\xiionst/\txn{erg}^{-1}\txn{Hz}) = 22.12 \pm 0.48 + 1.15 \pm 0.16 \times \log \big(\EW \,\OIIInoL / \txn{\AA}\big) \, ,
\end{equation}
while the grey bands indicate the 68 per cent credible region around the best-fitting relation.
We also fit a relation considering the combined equivalent widths of $\Hb + \OIII$, obtaining a relation with intercept (slope) $=21.95$ (1.18) instead of $22.12$ (1.15). Such a relation is particularly useful to convert in a physically-motived way broad-band colour excesses (e.g. IRAC excess, \citealt{Shim2011,Stark2013,Smit2014}) into \xiionst\ values.

\section{Discussion}\label{sec:discussion}

The results presented in the previous section allow us to draw several conclusions regarding  our modelling approach. The \beagle-based fitting of the UV and optical emission lines, and in particular our ability to match simultaneously the emission from different ionization states of the same element (e.g. $\txn{O}^+$ and $\txn{O}^{2+}$; $\txn{Ar}^{2+}$ and $\txn{Ar}^{3+}$), ions with widely different ionizing energies (e.g. 27.6 eV for $\txn{Ar}^{2+}$ and 41.0 eV for $\txn{Ne}^{3+}$) and electron temperature-sensitive ratios (e.g. $\OIIIc/\OIII$), suggests that the ionizing stellar-population prescription and nebular parametrization of our model are suited to interpret observations of extreme star-forming regions. The matching of the different H-Balmer lines validates our adoption of an SMC extinction curve to model dust attenuation, and that of the \SII\ line doublet the gas density we have adopted. Also, the ability of our model to match emission lines of different elements (C, N, O, Ne, S and Ar) suggests that these galaxies show similar metal abundance ratios, which we can reproduce by varying only the interstellar metallicity, dust depletion factor and carbon-to-oxygen ratio.  

We note, however, that in our analysis we did not attempt to fit the \HeIInoL\ recombination lines (\HeII\ and \HeIIopt). As we have discussed in S17, the stellar population + photoionization models incorporated in \beagle\ are unable to reproduce the very large equivalent widths of the \HeII\ nebular line (up to $\sim1.7$ \AA) observed in the galaxies of our sample that do not show WR spectral features, i.e. galaxies with metallicities $\logOH \lesssim 8$. As discussed in \citealt{Shirazi2012}, this problem is common to other widely used stellar population models, which fail in providing enough $\txn{He}^+$ ionizing photons for stellar populations with $\Z/\Zsun \lesssim 0.2$. Several potential sources of additional $\txn{He}^+$ ionizing photons have been proposed in the literature, such as X-ray binaries, fast radiative shocks, and low-luminosity AGNs \citep[e.g.][]{Stasinska2015}. As discussed in sec.~5.3 of S17, data on our galaxies do not provide evidence for the presence of X-ray binaries or fast radiative shocks, although deeper X-ray observations and larger grids of shock models (extending to lower metallicities) would be required to rule out such presence. Thanks to the recent incorporation into \beagle\ of the AGN models for Narrow Line Region emission of \citet{Feltre2016}, in a future work we will test the possibility of low-luminosity AGNs to explain the observed \HeII\ nebular emission of our non-WR galaxies.

The constraints obtained with our \beagle-based modelling of emission lines + photometry on \xiionst\ have potentially important implications for our understanding of the sources of H-ionizing photons in pristine star-forming galaxies. We find that \xiionst\ can reach values significantly larger than those considered in many reionization models, especially for galaxies exhibiting $\EW(\OIII) \gtrsim 1000$. Such galaxies are relatively rare at low redshifts, while they become much more common at higher ones. Very large equivalent widths of optical emission lines are routinely inferred from contamination of broad-band \Spitzer/IRAC photometry, corresponding to $\EW(\Ha+\NII+\SII)$ up to $\sim400$ for galaxies at $z\sim4.5$ \citep{Stark2013,Smit2016,Marmal2016}, reaching $\gtrsim1000$ for galaxies at $z\sim5.2$ \citep{Rasappu2016}. \citet{Smit2014} derive $\EW(\Hb+\OIII)\gtrsim1000$ for extremely blue IRAC galaxies at $z\sim7$. Observations of high-ionization UV lines provide a consistent picture: while high equivalent widths of UV lines, such as \CIV, \CIII\ and \OIIIuv, are rare at low redshift, they have been routinely detected at higher redshifts \citep[e.g.][]{Stark2015a, Stark2015b,Stark2017,Mainali2017}.
The emerging picture is that at $z\gtrsim6$ a combination of high star formation rates per unit stellar mass, young stellar populations, low metallicities and high ionization parameters favour the presence of $\EW\gtrsim1000$ (10) for the brightest optical (UV) lines, which our analysis suggest correspond to values of \xiionst\ significantly larger than canonical values used in reionization models.
 
From a theoretical perspective, simulations indicate that the escape fraction of H-ionizing photons from galaxies is time-dependent, being tightly coupled to the energy input from stellar feedback (winds, SNe), which opens `clear' channels through which ionizing radiation can escape \citep[e.g.][]{Wise2009, Trebitsch2017}. The presence of broad components in the emission lines of our objects suggests that such feedback mechanisms are at work in star-forming regions with a high production efficiency of H-ionizing photons. If regions with high \xiionst\ also exhibit large \fesc, then short, powerful bursts of star formation in low-metallicity, low-mass galaxies can provide enough H-ionizing photons to achieve the time-averaged value $\log (\fesc \, \xiionst/\mathrm{erg}^{-1}\mathrm{Hz}) = 24.5$ \citep{Bouwens2016} necessary to keep the IGM ionized.\footnote{This value assumes a minimum UV magnitude $M_{1500}=-13$, while adopting $M_{1500}=-17$ would imply a higher $\log (\fesc \, \xiionst/\mathrm{erg}^{-1}\mathrm{Hz}) = 24.9$.}  

\section{Comparison with previous work}\label{sec:comparison}

The importance of constraining the production rate of H-ionizing photons to understand the role of star-forming galaxies in cosmic reionization motivated several recent studies in which this quantity has been estimated using different observables. \citet{Bouwens2016} consider star-forming galaxies at \range{z}{3.8}{5.4} in the GOODS North and South fields, and estimate the production rate of H-ionizing photons \Nion\ from the contamination of $\Ha+\NII+\SII$ to broad-band photometry. In practice, they derive the \Ha\ luminosity by adopting fixed ratios of $\NIInoL/\Ha$ and $\SIInoL/\Ha$, and estimate the amount of dust attenuation from the UV-continuum slope. They then derive a dust-corrected UV luminosity \LuvHII. The \xiionHII\ they obtain span a broad range of values \range{\log(\xiionHII/\txn{erg}^{-1}\,\txn{Hz})}{24.2}{26}, compatible with the values spanned by our galaxies, with the bulk of their galaxies showing values \range{\log (\xiionHII/\txn{erg}^{-1}\,\txn{Hz})}{25}{25.6}, and the bluest and less luminous galaxies exhibiting the highest values.

\citet{Schaerer2016} analyse five LyC-leaking galaxies at $z \sim 0.3$ with large $\EW(\OIIInoL) \gtrsim 1000$, deriving \Nion\ from the (dust-corrected) \Hb\ luminosity, and \LuvHII\ from fitting the UV (GALEX) and optical (SDSS) photometry of the galaxies. As in \citet{Bouwens2016}, \citet{Schaerer2016} derive a dust-corrected UV luminosity, but without correcting for the contribution from recombination continuum and dust within \HII\ regions. The values they obtain are in the range \range{\log(\xiionHII/\txn{erg}^{-1}\,\txn{Hz})}{25.1}{25.5}, generally lower than the values we derive for galaxies with comparable $\EW(\OIIInoL)$. 

\citet{Matthee2017} study a sample of $\sim 600$ \Ha\ emitters at $z=2.2$ from the HiZELS survey \citep{Geach2008}. They measure the \Ha\ luminosity from a narrow photometric band centred on $\Ha+\NII$, correcting \Ha\ for the \NII\ contribution. They derive dust corrections with different methods, and compute \Nion\ from the dust-corrected \Ha\ luminosity. The \xiionHII\ they derive cover a broad range of values \range{\log(\xiionHII/\txn{erg}^{-1}\,\txn{Hz})}{24}{26}, with the exact values being sensitive to the adopted dust correction method. They find a strong correlation among \xiionHII\ and $\EW(\Ha)$, with the highest equivalent width galaxies exhibiting the large \xiionHII\ values.

\citet{Izotov2017a} study a sample of $\sim 14000$ compact star-forming galaxies at $z\le1$ from SDSS DR12. They derive dust corrections from Balmer line ratios, compute \Nion\ from the dust-corrected \Hb\ luminosity, and obtain the monochromatic UV luminosity from SED fitting to GALEX and SDSS photometry. They derive \xiionHII\ values in the range \range{\log(\xiionHII/\txn{erg}^{-1}\,\txn{Hz})}{24.4}{25.6}, with the bluest galaxies exhibiting the largest \xiionHII\ values.

\citet{Shivaei2018} consider a large sample of $\sim 700$ galaxies at redshift \range{z}{1.4}{2.6} from the MOSDEF survey \citep{Kriek2015}. They derive \Nion\ from the dust-corrected \Ha\ luminosity, and obtain \LuvHII\ by fitting the 3D-HST multi-band photometry with the \citet{Bruzual2003} population synthesis code. They apply average corrections to \Nion\ to account for a non-zero escape fraction of H-ionizing photons, and find $\log(\xiionHII/\txn{erg}^{-1}\,\txn{Hz})$ values in the range $\sim 25.00$ to $\sim 25.6$ (assuming an SMC extinction curve). As found by \citet{Bouwens2016} and \citet{Izotov2017a}, they also find that the bluest galaxies exhibit the largest \xiionHII\ values.
 
Measurements of the production rate of H-ionizing photons in EoR galaxies are difficult because of the challenges in performing spectroscopic observations of such galaxies with existing facilities. \citet{Stark2015b} and \citet{Stark2017} present detections of \Lya, \CIV\, and \CIII\ in 3 galaxies at $z\gtrsim7$, and estimate \xiionst\ from the stellar populations + photoionization modelling of their UV emission lines and photometry. They derive \range{\log(\xiionst/\txn{erg}^{-1}\,\txn{Hz})}{25.6}{25.8}, comparable to the largest values obtained in the present work.

\section{Conclusions}\label{sec:conclusions}

We have analysed a sample of ten extreme nearby star-forming regions with high-quality UV-to-optical spectra from \HST/COS, MMT and SDSS. Our objects are characterised by low metallicities and high ionization parameters, as suggested by the locus they occupe in the \NIInoL-BPT diagram and in the O32 \vs\ R23 one. 
%Unlike the objects of \citet{Izotov2017}, our galaxies do not reach extreme O32 values $\gtrsim10$, occupying in the O32 vs R23 diagram a locus similar to that occupied by the $z\sim0.3$ LyC leakers of \citet{Izotov2016} and by the $z\sim2.5$ star-forming galaxies of \citet{Strom2017}. 
Using a combination of stellar population + photoionization models incorporated into the \beagle\ spectral analysis tool, we have fitted a combination of UV-to-optical photometry + Balmer lines, and a broad set of UV and optical emission lines sensitive to the ionizing spectrum, ionization parameter, electrons temperature and density, metal abundances and  dust attenuation. This modelling has allowed us to constrain the intrinsic UV luminosity of the galaxies \Luvst, their production rate of H-ionizing photons, and their physical properties, such as gas-phase and stellar metallicities, ionization parameters, dust depletion factors, stellar ages and dust attenuation.

The ten galaxies of this work are characterised by large ionization parameters ($\logUs \gtrsim -3$), low dust depletion factors (\range{\xid}{0.1}{0.3}), low dust attenuation and very young stellar ages ($\t \lesssim 10 $ Myr in all but one case). As already shown in S17, our galaxies span a broad range of gas-phase metallicities \range{\logOH}{7.9}{8.5}. We have compared the \logOH\ values estimated with \beagle\ with those estimated in S17 with the direct-\Te\ method, finding an overall good agreement among the estimates, although the \beagle-based metallicities are systematically larger (by $\sim0.05$ dex) than the S17 ones. Larger differences ($\sim 0.2$ dex) among the two methods appear at $\logOH \gtrsim 8.3$. This confirms previous findings that metallicities derived with photoionization models are systematically larger than those derived with the direct-\Te\ method, although the differences found in this work are significantly smaller than those reported in the literature.

The production rate of H-ionizing photons per unit (intrinsic) UV luminosity of our galaxies spans a wide range [\range{\log(\xiionst/\txn{erg}^{-1}\,\txn{Hz})}{25.2}{25.8}] compatible with the values measured in previous works, but with a strong dependence of \xiionst\ on the equivalent width of the \OIII\ line, such that high-$\EW(\OIIInoL)$ galaxies exhibit larger \xiionst\ values. Such a relation is not completely surprising, since high-$\EW(\OIIInoL)$ are typically associated with young, metal-poor stellar populations, which are expected to emit large amounts of H-ionizing photons. Energy injection from young stellar populations (i.e. stellar `feedback') is also expected to create clear channels in the ISM through which ionizing radiation can escape, and it is likely the cause of the low-intensity broad components emission lines observed in our galaxies. The combination of large \xiionst\ and \fesc\ in short bursts of star formation in low-mass galaxies can hence supply large quantities of H-ionizing photons to the IGM. To enable more accurate determinations of \xiionst\ in EoR galaxies, we have calibrated a new relation between \xiionst\ and $\EW \, \OIII$ ($\EW \, \OIII + \EW \, \Hb$). This relation can be used to derive \xiionst\ from broad-band observations of galaxies at z$\gtrsim6$, e.g. from existing \Spitzer/IRAC observations and future \JWST/NIRCam ones, and from spectroscopic observations of \OIII\ with \JWST/NIRSpec out to redshift $\sim9.5$.

\section*{Acknowledgements}
 
The authors are grateful to the referee, G. Stasi\'nska, for several comments that significantly improved the paper. JC, SC, and AVG acknowledge support from the European Research Council (ERC) via an Advanced Grant under grant agreement no. 321323-NEOGAL. DPS acknowledges support from the National Science Foundation through the grant AST-1410155. TJ acknowledges support from NASA through Hubble Fellowship grant HST-HF2-51359.001-A awarded by the Space Telescope Science Institute. AF acknowledges support from the ERC via an Advanced Grant under grant agreement no. 339659-MUSICOS. The data described in this paper were obtained by the MMT Observatory, a joint facility of the University of Arizona and the Smithsonian Institution, and the NASA/ESA Hubble Space Telescope.  Support for HST GO program \#14168 was provided by NASA through a grant from the Space Telescope Science Institute, which is operated by the Association of Universities for Research in Astronomy, Inc., diunder NASA contract NAS\,5-26555.

\bibliographystyle{mnras}

\bibliography{Local_analogues_Cycle_23} % your references Yourfile.bib

%TC:endignore

% Don't change these lines
\bsp	% typesetting comment
\label{lastpage}
\end{document}